# Evidence for the evolution and decay of an electrified Medium Scale Traveling Ionospheric Disturbances during two consecutive substorms: First results

R. Rathi[1], M. Sivakandan[2], D. Chakrabarty[3], M. V. Sunil Krishna[1,4], A. K. Upadhayaya[5], S. Sarkhel[1,4*]

[*]Sumanta Sarkhel, Department of Physics, Indian Institute of Technology Roorkee, Roorkee - 247667, Uttarakhand, India (sarkhel@ph.iitr.ac.in)

[1]Department of Physics,
Indian Institute of Technology Roorkee,
Roorkee - 247667
Uttarakhand, India

[2]Leibniz Institute of Atmospheric Physics (IAP),
Kühlungsborn,
Germany

[3]Space and Atmospheric Sciences Division,
Physical Research laboratory,
Ahmedabad-380009
Gujarat, India

[4]Centre for Space Science and Technology, Indian Institute of Technology Roorkee, Roorkee – 247667, Uttarakhand, India

[5]Environmental Sciences and Biomedical Metrology Division,
CSIR National Physical Laboratory, New Delhi, India

**Abstract**

Electrified Medium Scale Traveling Ionospheric Disturbances (EMSTIDs) is one of the prominent plasma structures that affect the propagation of high frequency radio waves. Overall, seasonal variation and propagation characteristics of the EMSTIDs are widely reported in literature. However, the effects of substorms on the formation and dissipation of the EMSTIDs are not well explored. In the present study, on a moderately geomagnetically active night of 26 October 2019 (Ap=24), the airglow imager over Hanle (32.7°N, 78.9°E; Mlat. ~24.1°N), India recorded the evolution and decay of an EMSTID in the O($^1$D) 630.0 nm airglow images in between 13.3 UT and 15.8 UT. In addition, during the same time, a steep rise and fall of the virtual base height of the ionospheric F-layer were also recorded by a nearby digisonde over New Delhi (28.70°N, 77.10°E; Mlat. ~20.2°N). The most important aspect of the event was the occurrence of the two consecutive substorms in between 13.3 UT and 15.8 UT. To the best of our knowledge, this is the first of its kind study where we report the role of interplanetary electric field (IEF) and substorm induced electric fields on the evolution and decay of the EMSTID. This study elicits effects of the externally imposed electric fields on the mid-latitude ionospheric plasma structures and provides insight into the complex coupling between auroral and low-mid latitude region.

**Highlights:**

1. Evolution and dissipation of an EMSTID were recorded in the O($^1$D) 630.0 nm all-sky airglow images.
2. Eastward and westward prompt electric fields were observed over Hanle during two consecutive substorms.
3. Interplanetary and substorm induced electric fields can significantly affect the mid-latitude ionosphere and plasma structures.

## 1. Introduction

Electrified medium scale traveling ionospheric disturbances (EMSTIDs) in which the phase fronts are aligned in the northwest to southeast (northeast to southwest) and propagate south-westward (north-westward) in the northern (southern) hemisphere. Most often EMSTIDs occurs during solstice in the ionospheric mid-latitude region (Ding et al., 2011; Huang et al., 2016; Kotake et al., 2006; Martinis et al., 2010; Shiokawa et al., 2003a and references therein). Perkins instability is believed to be a causative mechanism for the generation of EMSTIDs (Garcia et al., 2000; Hamza, 1999; Perkins, 1973). However, investigations showed that the linear growth rate of the Perkins instability is very small. On the other hand, Perkins instability in the combination with some other processes such as E-F coupling, gravity waves, and interhemispheric coupling could explain the growth of the EMSTIDs (Cosgrove and Tsunoda, 2004; Huang et al., 1994; Kelley & Makela, 2001; Narayanan et al., 2018; Perkins, 1973; Otsuka et al., 2004; Shiokawa et al., 2005; Sivakandan et al., 2022). These studies suggested that the polarization electric field can map along equipotential geomagnetic field lines from E region and/or geomagnetic conjugate F-region. This polarization electric field provides positive feedback to the Perkins instability, enhances its growth rate, and hence generates EMSTIDs. Along with the evolution, dissipation/disappearance of the EMSTIDs have been studied in the recent times. In literature, there are very few studies which have reported the dissipation/disappearance of EMSTIDs and its probable causative mechanism (Luo et al., 2021; Narayanan et al., 2014; Rathi et al., 2022; Sivakandan et al., 2019; Wu et al., 2021). EMSTIDs may disappear when they encounter the equatorial ionization anomaly (EIA) crest due to the enhanced electron density (Narayanan et al., 2014; Sivakandan et al., 2019). Further Wu et al. (2021) and Rathi et al. (2022) suggested that the stronger resultant electric field resulted after the interaction of EMSTIDs' fronts contributed to their dissipation, whereas Luo et al. (2021) suggested that the increased equatorward wind can also play a role in the dissipation of the EMSTIDs. These studies investigated the role of polarization electric field and ionospheric background conditions on the generation and dissipation of the EMSTIDs. However, the role of induced electric fields due to geomagnetic disturbances (storms and/or substorms) on the various aspects of the EMSTIDs' evolution is not well understood.

A geomagnetic storm is defined as depression in the horizontal component (H) of the geomagnetic field due to the intensification of the ring current caused by solar events such as Coronal Mass Ejection (CME) or Co-rotating Interaction Region (CIR) (Gonzales et al., 1994; and references therein). The depression in the H is represented by Disturbance storm time (Dst) index, which defines the magnitude of the geomagnetic storms (Gonzales et al., 1994; Kamide et al., 1998). On the other hand, a substorm can be defined as a process of storage of energy in the magnetosphere and its subsequent release in an explosive manner. Substorms, in general, are also associated with the formation of partial ring current and intensification of the auroral electrojet (Gonzales et al., 1994). Substorms can be identified by many features such as dispersion-less injection of energetic electrons and ions in the geosynchronous orbit, intensification of westward auroral electrojet (AL & SML index depression), increase in the polar cap index (PC), magnetic bay signatures, longitudinally confined substorm current wedge formation etc. (Chakrabarty et al., 2008; Huang, 2005; McPherron and Chu, 2018; Newell & Gjerloev, 2011; Reeves et al., 1996; Saito et al., 1976; Troshichev et al., 1979; Weygand et al., 2008). Similar to geomagnetic storms, substorms can also cause global ionospheric effects although substorms occur in a localized manner (Belehaki et al., 1998; Rout et al., 2019). Various studies reported the substorm induced prompt electric fields over equatorial and mid-latitude region (Chakrabarty et al., 2010 & 2015; Fejer et al., 2021; Hui et al., 2017; Kikuchi et al., 2000 & 2003; Kumar et al., 2023; Park, 1971; Reddy et al., 1990; Rout et al., 2019; Veenadhari et al., 2019). Chakrabarty et al. 2010 and Rout et al. 2019 reported the substorm induced eastward electric field over equatorial region. Numerous studies showed that substorms can cause westward electric field perturbations (Chakrabarty et al., 2015; Kikuchi et al., 2000 & 2003; Veenadhari et al., 2019). Whereas, in their studies, Hui at al. 2017 and Fejer et al. 2021 reported both eastward and westward electric field perturbations over equatorial region. In addition, a few studies showed that the substorm can induce electric fields over mid-latitude ionosphere (Park, 1971; Reddy et al., 1990). Park, 1971 reported the presence of westward electric field during substorm. Whereas, the investigation led by Reddy et al. 1990 showed that both eastward and westward electric fields can be present over the mid-latitude region during substorms. These studies suggested that substorm can generate both eastward and westward electric field perturbations. The effects of these electric field perturbations such as variations in the vertical plasma drift and 630.0 nm airglow intensity in the low and mid-latitude ionosphere have been investigated extensively. However, the effects of substorm related prompt electric field on the mid-latitude plasma structures (such as EMSTIDs) are not well understood.

Over the four years of observation using Hanle imager, there are a few cases in which multiple interesting events are reported. For example, Sivakandan et al. (2020, 2021) reported mid-latitude spread F (MSF) and in-situ generated plasma depletion. Yadav et al. (2021a, 2021b) investigated complex interactions of EMSTIDs and field aligned plasma depletion. In their studies, Rathi et al. (2021, 2024a) and Patgiri et al. (2024a) reported the existence of various plasma structures (like periodic EMSTID, single dark band EMSTID, and gravity wave MSTID) having different characteristics whereas, Chakrabarti et al. (2024) found a novel deep learning technique for automatic detection of plasma structures in the all-sky images. In another study, Rathi et al. (2022) reported bifurcation and dissipation of the EMSTIDs during the interaction of the two fronts. Recently, Patgiri et al. (2024b) investigated the modifications in the MSTID's bands resulted due to the multiple self-interactions. It is to be noted that all these events were observed during geomagnetic quiet conditions (Kp<3). Most of the earlier investigations reported the EMSITDs occurrence during the geomagnetic quiet condition (Ding et al., 2011; Huang et al., 2016; Sun et al., 2015) and/or storm events (Nishioka et al., 2009; Terra et al., 2020). In the literature, the role of substorm on the evolution, propagation characteristics and/or dissipation of the EMSTIDs are not well reported so far. In the present communication, we report a unique set of imaging observations that showcase the influence of $Y$-component of the interplanetary electric field (IEF$_y$) and substorm induced electric fields on the evolution and decay of an EMSTID.

## 2. Instrument and Data Analyses

We have used 630.0 nm airglow data from a multi-wavelength all-sky airglow imager installed at Hanle, Ladakh, India (32.7°N, 78.9°E; Mlat. ~24.1°) which lies beyond the northern edge of the equatorial ionization anomaly crest region. Hanle imager consists of a highly-sensitive thermoelectrically cooled CCD camera with a fisheye lens and filter wheel, which captures images of airglow emissions at 630.0 nm (centered at 250 km) and 557.7 nm (centered at 97 km) wavelengths. In order to get the information of each pixel (such as azimuthal, elevation, and geographic coordinates) in the raw images, we do geospatial calibration by identifying some bright stars. After geospatial calibration, we remove CCD dark, sky, and star noises. Afterwards, unwarped images are produced by using equidistant gridding for each pixel and use these images for analyses. For more detailed descriptions of the imager and image processing techniques, we refer the readers to Mondal et al. (2019).

Apart from the all-sky imager, ionograms of 15-minute resolution obtained from New Delhi digisonde (28.70°N, 77.10°E; Mlat. ~20.2°N) are used as complementary dataset. We scaled the F-layer virtual base height (h'F) and examined the presence of spread F and sporadic E layer by using the ionograms. A more detailed description about the digisonde can be found in Upadhayaya and Mahajan (2013). In addition, we have also used geomagnetic field data from ground magnetometers obtained from SuperMAG.

In addition to the ground-based observations, we have also used the interplanetary electric field and magnetic field from the Advanced Composition Explorer (ACE) satellite obtained from NASA GSFC CDAWeb. The *Y*-component of interplanetary electric field (IEF$_y$) is derived based on the solar wind velocity ($V_{sw}$) and *z* component of interplanetary magnetic field (IMF $B_z$) data using the equation:

$$IEF_y = V_{sw} \times B_z \qquad (1)$$

It is to be noted that the data obtained from NASA GSFC CDAWeb is already corrected for time lag right up to the nose of the terrestrial bow shock. Later, following the methodology of Chakrabarty et al. (2005), we calculated the magnetosheath and Alfven transit times and added to the lag time. In the present study, we used time lag corrected data of IEF$_y$ and IMF $B_z$. In addition, we have also used energetic electron flux data at four different energy channels (75, 150, 275, and 475 keV) measured by Geostationary Operational Environmental Satellite system (GOES)-15.

## 3. Results

### *3.1 Evolution and Decay of an EMSTID over Hanle*

An EMSTID event was observed in the O($^1$D) 630.0 nm airglow images on 26 October 2019 over Hanle (32.7°N, 78.9°E; Mlat. ~24.1°N), Ladakh, India. We can observe that very faint EMSTID's fronts started to appear around 13.31 UT or 18.81 IST (Indian Standard Time = UT + 5.5 h), which became prominent later and present in the field of view (FOV) of the imager till 15.81 UT (21.31 IST). A sequence of unwarped images showing this entire event is presented in Figure 1. In each subfigure, the location of the airglow imager at Hanle and digisonde at New Delhi is indicated by a red star and triangle, respectively and the green dotted line indicates the geomagnetic field line passing over Hanle. We have provided the entire sequence of unwarped images in the form of a movie available in the supplementary material for reader's convenience. The overall intensity of the airglow images was low during the interval of 13.52-14.56 UT (19.02-20.06 IST) (Figures 1b-e). In the beginning EMSTID's front are not observed in the FOV. A very faint dark (low airglow intensity) front with northwest-southeast (NW-SE) orientation marked with yellow dashed-dotted line appeared in the images around 13.31 UT (18.81 IST) (shown in Figure 1a). The dark front evolved with time (after 13.52 UT or 19.02 IST) and became more prominent (darker) in later hours (Figures 1b-g). Along with the first front, another dark front marked with magenta dashed-dotted line also appeared and evolved (Figures 1d-g). Later, the overall intensity of the airglow images started to increase around 14.56 UT (20.06 IST) (Figures 1f-l). Furthermore, the process of evolution of the EMSTID's fronts reversed to decay (fronts became fainter) from around 14.87 UT (20.37 IST) and the fronts completely disappeared (Figures 1h-o) nearly at 15.81 UT (21.31 IST). A notable aspect of this event is that the EMSTID evolved and decayed within the FOV of the imager with a time span of two and a half hours. In addition to the evolution and decay of the EMSTID's fronts, one more NW-SE aligned very faint small dark front appeared in the southern part of the bright band of the EMSTID around 14.87 UT (20.37 IST) as marked by cyan dashed-dotted line, which later split into thin fronts marked as 'd$_1$', 'd$_2$', and 'd$_3$' (Figures 1h-l). In this event, we observed multiple intriguing features viz. temporal changes (reduction and enhancement) in the overall 630.0 nm airglow images' intensity, evolution and decay of an EMSTID, and splitting of a MSTID front.

### *3.2 Variation in the IEF$_y$, virtual base height (h'F) of the F-layer, vertical E × B drift, amplitude of the EMSTID, and ΔX of the geomagnetic field*

On occasions, MSTID propagation may cause undulations in the F-layer height as a result spread in the height and/or frequency can occur in the ionograms (Amorim et al., 2011; Narayanan et al., 2018; Shiokawa et al., 2003a). In order to check whether any undulation in the F-layer occurred or not during this event, we investigated the ionogram data using New Delhi digisonde. Figure 2 shows a sequence of the ionograms on the night of 26 October 2019. An ascending and descending in the altitude of the F-layer traces were observed during the intervals of 13.5-14.5 UT (19.0-20.0 IST) and 14.5-15.5 UT (20.0-21.0 IST), respectively. In addition to that few ionograms show a weak spread-F between 14.5 and 16.75 UT (20.0 and 22.25 IST). In order to check any role of Y-component of interplanetary electric field (IEFy) on the F-layer height variation, we have plotted the temporal variation of the time lag (due to delays between ACE satellite observation at L1 point and the ionospheric observation) corrected IEF$_y$ (black curve), along with IMF B$_z$ (green curve) (Figure 3a). IEF$_y$ was positive (duskward) during the first substorm and became negative (dawnward) at the end of the first substorm. It again became positive (duskward) after the first substorm till ~14.9 UT (20.4 IST). According to the classic scenario in the dayside (nightside) a duskward (dawnward) IEF$_y$ is associated with the eastward (westward) PPEF in the ionosphere (Wei et al., 2015). However, there are some studies which observed and reported the duskward IEF$_y$ (or eastward PPEF in the ionosphere) till 22:00 LT. The modelling (e.g., Nopper & Carovillano, 1978) and observational (e.g., Fejer et al., 2008; Kumar et al., 2023) studies suggested that the eastward perturbations of PPEF is expected till 22:00 LT. Therefore, observation and interpretations in the present study are consistent with the previously existing understandings. Also, using the ionograms, the virtual base height (h'F) of the F-layer has been estimated. Figure 3b depicts the temporal variation of the h'F on the night of 26 October 2019 (black curve) and average h'F of the quiet nights in the month of October 2019 (cyan curve). The presence of weak spread-F and blanketing sporadic E (Esb) layers is marked by green stars and magenta asterisks, respectively. The quiet time averaged h'F in October 2019 showed a gradual increment after 14 UT (19.5 IST) whereas on 26 October 2019, it showed a steep rise (13.52-14.5 UT or 19.02-20.0 IST) and fall (14.5-15.5 UT or 20.0-21.0 IST) of ~70 km between 13.5 and 15.5 UT. Following a method by Nishioka et al (2009), we have also calculated the $\boldsymbol{E} \times \boldsymbol{B}$ drift using h'F on 26 October 2019. First, we calculated the apparent vertical drift (dh'F/dt) by dividing the difference in successive h'F (dh'F) to their time difference (dt). This drift included all the contributing factors viz $\boldsymbol{E} \times \boldsymbol{B}$ drift, chemical loss [βH, where β is effective chemical loss coefficient and H is plasma scale height/ionization scale height calculated using the measured electron density profiles (following Devasia et al., 2001; Krishna Murthy et al., 1990; Sreekumar & Sripathi, 2016)], and neutral winds effect [$U(cosI)(sinI)$, where $U$ is meridional wind, and $I$ is the inclination angle]. The effective chemical loss co-efficient and plasma scale height are calculated using NRL-MSISE00 model (Picone et al., 2002) and ionosonde data, whereas the meridional winds are obtained from HWM-14 (Drob et al., 2015). In order to find the $\boldsymbol{E} \times \boldsymbol{B}$ drift ($\boldsymbol{v}$), the contribution of βH and neutral wind are subtracted from the apparent vertical drift. Figure 3b shows the variation of $\boldsymbol{E} \times \boldsymbol{B}$ drift (red curve) between 13.0 and 15.75 UT (18.5 and 21.25 IST). There was an upward $\boldsymbol{E} \times \boldsymbol{B}$ drift between 14 and 14.5 UT (19.5 and 20.0 IST) and downward $\boldsymbol{E} \times \boldsymbol{B}$ drift between 14.5 and 15.75 UT (20.0 and 21.25 IST). It is to be noted that the magnitude of the estimated drift is larger than the maximum uncertainty (~10 m/s) calculated by Bullet (1994) and woodman et al. (2006). This supports that the polarity of the $\boldsymbol{E} \times \boldsymbol{B}$ drift during the two substorms is reliable. Temporal variation of the foF2 is also shown in Figure 3b (gray curve), which showed similar trend with F-layer height till 15 UT (20.5 IST), whereas it showed anomalous behavior between 15 and 15.5 UT (20.5 and 21.0 IST) during second substorm. It increased even when the F-layer descended. This anomalous enhancement in foF2 was probably due to the passage of an enhanced plasma density region over the New Delhi digisonde (as shown in the sequence of global VTEC in Figure S2 of supplementary document). Figure 3c shows the temporal variation of the intensity of the EMSTID (intensity of the EMSTID) without removal of the background intensity (magenta curve), amplitude of the EMSTD after removing background airglow intensity (blue curve), and change in the north-south component (ΔX) of the geomagnetic field (black curves). At first, the intensity of the first dark front of the EMSTID (intensity of the EMSTID) is estimated by taking the minimum intensity along a line perpendicular to the EMSTID front in each image (shown in Figure S1 of the supplementary document). This shows significant decrease between 13.5 and 14.3 UT (19.0 and 19.8 IST) and after that it started increasing (shown by magenta curve in Figure 3c). It is to be noted that, during this time, the overall intensity of the airglow images also varied. Therefore, the intensity of the EMSTID without removing the background airglow intensity has a dominant contribution from the background itself. Thus, we determined the amplitude of the EMSTID (excluding background intensity) i.e., maximum percentage deviation corresponding to EMSTID's front. In order to find the amplitude, we followed the methodology reported in Huang et al. (2016) and Shiokawa et al. (2003a), in which they determined the percentage deviation [$\Delta I_D = \left(\frac{I-I_o}{I_o}\right) \times 100 (\%)$; where $I$ and $I_o$ are the observed airglow intensity and the running averaged intensity respectively]. In the present event, we subtracted the 400 km running mean of the intensity ($I_o$) from the intensity ($I$) extracted along a line perpendicular to the EMSTID front, the obtained residual contains the perturbation below 400 km. This analysis was carried out for each image until the dissipation of the EMSTID. Using these residual intensities, we determined the percentage deviation ($\Delta I_D$) for each image. Using these percentage deviations, we determined amplitude (i.e., maximum percentage deviation) corresponding to the first dark front of the EMSTID. We tracked the first dark front, in which the background airglow intensity eliminated and plotted its temporal variation (shown in Figure 3c blue curve). It is clear that the intensity of the EMSTID (without removing the background) decreased during the first substorm whereas it increased during the second substorm. However, the amplitude of the EMSTID after removing the background airglow intensity started increasing after the first substorm onset, whereas it started decreasing with a slight time delay after the second substorm onset. This confirms that the overall intensity and amplitude of the EMSTID behaved in an opposite way. Figure 3c also shows the temporal variation $\Delta X$ from three ground magnetometers [Novosibirsk (55.03°N, 82.9°E), Alma Ata (43.25°N, 76.92°E), and Alibag (18.62°N, 72.87°E)] near to the Hanle longitude (78.9°E). It is clear from the figure that $\Delta X$ shows characteristically similar variations during both the substorm intervals. During the first substorm, $\Delta X$ showed minor concomitant decrease over all the three stations, whereas during the second substorm, it showed a remarkable nearly simultaneous enhancement over all the three stations (Figure 3c). These results suggest that the variations in the magnetic field during these intervals are of magnetospheric origin. This argument is also supported by the fact that the local ionospheric conductivity during local post-sunset/nighttime is very low to support such current. This aspect will not be discussed further. In addition, the variation in the ΔX and dawnward turning of IEF$_y$ (or northward turning of IMF B$_z$) pointed towards the occurrence of substorms. Therefore, we also checked the occurrence of the substorms using various substorm associated features. We have marked the periods of the substorms by blue and red shaded regions in Figure 3. The detailed description about the substorms have been included in the following subsection.

*3.3 Two consecutive substorm events*

Looking at the Ap index (Ap=24), we found a moderately geomagnetically active condition during this event. We also checked for the substorm activity on this night. A substorm can be identified by numerous substorm associated features such as dispersion-less particle injections at the geosynchronous/geostationary orbits, depressions in AL or SML indices, increase in PC index, increase in asymmetric component of ring current (ASY-H) index, formation of magnetic bay signatures, enhancement in wave and planetary (Wp) index (which reflects Pi 2 activities at low and mid-latitudes), etc. (Chakrabarty et al., 2008; Huang, 2005; Li et al., 2009; Li et al., 1998; McPherron and Chu, 2018; Newell & Gjerloev, 2011; Nosé et al., 2012; Reeves et al., 1996; Saito et al., 1976; Troshichev et al., 1979; Weygand et al., 2008; and references therein) etc. Figure 4 depicts the temporal variation of different geomagnetic indices namely AL, PC, ASY-H, SYM-H, SMU, SML, Wp, and electron fluxes at four energies (75, 150, 275, and 475 keV) on 26 October 2019. A sharp decrease in the AL at ~13.52 UT (19.02 IST) and ~14.87 UT (20.37 IST) represents the expansion phase of the first and second substorm, which shows the onset of the two substorms, respectively (Figure 4a). During these two substorms, magnitude of the AL index reached below -1000 nT with a first minima (around -1100 nT) at ~13.7 UT (19.2 IST) and second minima (around -1600 nT) at ~15.2 UT (20.7 IST). The consecutive enhancements in PC index, ASY-H, SYM-H, variation in the SMU/SML indices (Figures 4a, c, e), and enhancements in GOES electron fluxes nearly from 13.52 UT (19.02 IST) and 14.87 UT (20.37 IST) (Figures 4b, d, f, h) confirmed the onset of the two substorms. In addition, the significant enhancement in the wave and planetary (Wp) index between 13.5 and 15.8 UT also confirmed the substorm onsets (Figure 4g). The response (variation) time of all these indices is consistent with the onset of the two substorms. The first substorm occurred during the interval of 13.52-14.3 UT (19.02-19.8 IST) and the second one occurred during 14.87-15.8 UT (20.37-21.3 IST) (Figures 4a-h). The blue and red dashed region in all the subfigures represent the duration of the first and second substorm, respectively.

## 4. Discussion

Nighttime EMSTIDs, one of the prominent plasma structures of the mid-latitude ionospheric F-region, have been extensively studied in the past few decades. Most of the earlier studies reported the occurrence of the EMSTIDs during the geomagnetic quiet condition (Ding et al., 2011; Huang et al., 2016; Sun et al., 2015) and a very few studies during the geomagnetic storm conditions (Nishioka et al., 2009; Terra et al., 2020). However, there is no extensive study on the role of substorm (which occurs more frequent than the geomagnetic storms) on the evolution and dissipation of the EMSTIDs. One probable reason for the lack of studies could be the effects of substorm are highly localized. In this present study, we have investigated the impact of the substorm on the evolution and dissipation of the EMSTID over a geomagnetic low-mid latitude transition region in India.

The event observed by the Hanle imager on the night of 26 October 2019 with moderate geomagnetic activity (Ap=24) covered the evolution as well as the decay of the EMSTID within two and a half hours. As already mentioned in *section 3.1*, the evolution of the EMSTID was started after 13.52 UT (19.02 IST) and completely dissipated at 15.81 UT (21.31 IST). Along with the evolution and decay of EMSTID, the overall intensity of the O($^1$D) 630.0 nm airglow images also showed large variation. For example, the airglow images became darker (intensity decrement) between 13.52 and 14.56 UT (19.02 and 20.06 IST) (Figures 1b-e) followed by the brightening (intensity increment) of the airglow images during the interval of 14.56-15.6 UT (20.06-21.1 IST) (Figures 1f-j). However, due to the presence of the EMSTID fronts in the whole FOV, it is not possible to find the background intensity quantitatively on 26 October 2019. During the decay phase of the EMSTID, a few small northwest to southeast aligned southwestward moving dark fronts were also observed in the southern part of the images. In this section, we will first discuss the plausible reasons behind the intensity variation and then about the evolution and decay of the EMSTID.

### *4.1 Cause of airglow intensity variation*

The O($^1$D) 630.0 nm airglow emission during nighttime arises in the 250 ± 40 km altitude range due to dissociative recombination mechanism (Link & Cogger, 1988):

$$O^+ + O_2 \rightarrow O_2^+ + O \qquad (2)$$
$$O_2^+ + e^- \rightarrow O^* + O \qquad (3)$$

Here, O$^*$ is the excited state, which may leave either in the O($^1$D) and/or O($^1$S) state. It is clear from the above equations (2) and (3) that O($^1$D) 630.0 nm airglow emission intensity depends on the electron, O$^+$, and O$_2$ densities. Therefore, the variations in the emission intensity (increment/decrement) could be caused by the F-layer height variation (fall/rise) and/or electron density variation (increment/decrement) (Makela et al., 2001; Makela & Kelley, 2003). In order to examine the observed intensity variations, we checked the variations in the virtual base height (h'F) of the F-layer over digisonde at New Delhi (a nearby station to Hanle imager). A steep rise and fall in the h'F was recorded during 13.52-14.5 UT (19.02-20.0 IST) and 14.5-15.5 UT (20.0-21.0 IST), respectively (Figure 3b). In addition, the peak frequency of the F2-layer (foF2) also showed nearly similar trend with F-layer height till 15 UT (20.5 IST), whereas it showed anomalous behavior between 15 and 15.5 UT (20.5 and 21.0 IST) during second substorm due to the transport of enhanced plasma density from eastern region (as shown in the sequence of global VTEC in Figure S2 of supplementary document). Using the h'F variations during this period, we had calculated the vertical $\boldsymbol{E}\times\boldsymbol{B}$ drift by subtracting the other contributing factors viz. βH and neutral wind contribution. The vertical upward and downward $\boldsymbol{E}\times\boldsymbol{B}$ drift were observed between 14-14.5 UT (19.5-20.0 IST) and 14.5-15.5 UT (20.0-21.0 IST), respectively (Figure 3b). These results confirm that the observed decrement and increment in the intensity of the 630.0 nm airglow emission was caused by the enhancement and reduction in the h'F due to the upward and downward $\boldsymbol{E}\times\boldsymbol{B}$ drift, respectively. The ascent/descent of F-layer, decrease/increase in 630.0 nm intensity, and upward/downward $\boldsymbol{E}\times\boldsymbol{B}$ drifts confirmed the presence of eastward/westward electric fields.

In order to find the source of these electric field perturbations, we also examined the role of the interplanetary electric field (IEF) and the two consecutive substorms occurred during this period. The first substorm occurred between 13.52 and 14.3 UT (19.02 and 19.8 IST), while the second occurred between 14.87 and 15.8 UT (20.37 and 21.3 IST) (Figure 4). The occurrences of these substorms are confirmed by the sharp decrease in the AL index, dispersion-less electron injections at the geosynchronous orbit, enhancement in PC, ASY-H, SYM-H, and Wp indices (Figure 4). During substorms, a rapid depolarization of geomagnetic field leads to the generation of induced electric fields in the magnetosphere (Ilie et al., 2017; Mishin et al., 2015; Pellinen et al., 1984), which causes electric field disturbances in the ionosphere also known as substorm induced electric fields. These electric fields cause ascent and/or descent of the ionospheric F-layer (Chakrabarty et al., 2008, 2010 & 2015; Rout et al., 2019; Fejer & Navarro, 2022) similar to the PPEF associated with the $IEF_y$ polarity reversal. In the present study, the ascent and descent of the F-layer was observed during two consecutive substorms and polarity reversal of $IEF_y$, therefore it is crucial to investigate the role of both these factors. It is to be noted that during the first substorm *Y*- component of interplanetary electric field ($IEF_y$) was duskward (positive) (Figure 3a). However, it is shown that 7-10% $IEF_y$ penetrates to the equatorial/low latitude ionosphere (Huang et al., 2007; Kelley et al., 2003; Nicolls et al., 2007). Therefore, considering 10% penetration efficiency of the duskward IEF$_y$ during the first substorm causes nearly ~11 m/s upward drift, which is comparatively lesser than the upward drift (~32 m/s) obtained. This suggests that, in addition to the eastward PPEF due to duskward IEF$_y$, other factors (induced electric field by the substorm and the equatorward wind) could have played role in the observed upward drift during first substorm. There might be two possibilities for the eastward electric field perturbation over Hanle. Both prompt penetration of IEF$_y$ and the substorm induced prompt electric field were eastward and contributed to the net eastward electric field perturbation. Another possibility is that the substorm induced prompt electric field was westward and the eastward prompt penetrated IEF$_y$ overpowered the substorm induced prompt electric fields generating a net eastward electric field perturbation. Since the present study is during a minor geomagnetic storm (Dst between -30 nT & -50 nT, and SYM-H was down to -50 nT) and two consecutive substorms (AL: ~ -1100 nT & -1600 nT), therefore we expected equatorward wind during both the substorms and could not neglect it's role, which, along with the eastward electric field caused the upward drift over the observation region. In the present study, during the first substorm two factors viz. upward $E \times B$ drift by eastward electric field (due to eastward PPEF and/or induced by the first substorm) and upward drift by equatorward wind combinedly played a role in the enhancement of the h'F, which consequently caused the reduction in 630.0 nm airglow images' background intensity. After the first substorm, when the effect of eastward imposed electric field vanished, the F-layer started descending at ~14.5 UT (20.0 IST). It is interesting to note that nearly after 14.87 UT (20.37 IST) the rate of the descent of the F-layer enhanced (also evident form the increased downward $E \times B$ drift) even though the wind was expected to be equatorward (Figures 3a, b). Simultaneously, we observed the duskward turning of IEF$_y$ at ~14.87 UT (20.37 IST), which triggered the second substorm. During the second substorm, westward prompt electric field perturbations caused by the substorm overpowered the equatorward wind. This westward electric field caused the rapid decrement in the h'F. It might be possible that the decrement in the h'F was the normal decrement after the first substorm (which caused the upliftment of the F-layer). However, during this time the decrease in the h'F was steep and more than 1σ value of the quiet time average h'F for the October month (Figure 3b). In addition, it dipped below the quiet time monthly averaged background h'F (Figure 3b). Therefore, it is reasonable to infer that the additional decrement in the h'F was due to the westward prompt electric field perturbations induced by the second substorm, which caused a downward $E \times B$ drift. The foF2 also showed increment during the second substorm, which resulted from the enhanced plasma density over the observation region (shown in Figure S2 of supplementary document). Thus, the decrease in the h'F due to downward $\mathbf{E}\times\mathbf{B}$ drift and enhanced plasma density during second substorm increased the overall 630.0 nm airglow intensity. These results support the presence of external eastward and westward electric fields over the Hanle region during these substorms. The present study also shows the simultaneous presence of the substorms augmented and/or annulled PPEF due to the IEF$_y$ as reported in a few earlier studies (Hui et al., 2017; Rout et al., 2019).

*4.2 Substorm impact on the evolution and dissipation of the EMSTIDs*

The EMSTID onset was observed at 13.31 UT (18.81 IST) with a very faint dark front, which evolved with time (after 13.52 UT or 19.02 IST) and became more prominent in later hours (Figures 1b-e). Along with it, a few more fronts were also evolved with time (Figures 1d-e), which later started decaying at after 14.87 UT (20.37 IST) and completely disappeared at 15.81 UT (21.31 IST) (Figures 1h-o). The evolution and decay of the EMSTID is also evident from the reduction and enhancement of the intensity of the EMSTID front (magenta curve in Figure 3c). In addition, the amplitude (maximum percentage deviation) of the EMSTID showed enhancement after first substorm onset and the reduction during the second substorm with a slight time delay (shown by blue curve in Figure 3c). In this observation, we can detect a prompt response of electric field changes on the h'F and background airglow intensity after the onset of both the substorms (Figures 3b & c). However, during the second substorm, the slight delay in the variation of the EMSTID's amplitude was due to the interplay between the polarization electric field and the external westward electric field. In other words, the polarization electric field that got enhanced due to the positive feedback provided by the eastward electric field (during the first substorm) could be significant enough to survive against the external westward electric field, which later became dominant and led to the reduction in the amplitude of the EMSTID. However, the increment rate of the amplitude of the MSTID started reducing after the onset of the second substorm upto 15.2 UT (shown by blue curve in Figure 3c). Thus, the reduction in the intensity and increase in the amplitude of the EMSTID indicated the evolution of the EMSTID, whereas the increase in its intensity and reduction in the amplitude indicated the decay of the EMSTID. It is important to note that the fronts of the EMSTID didn't propagate from the NE region of the Hanle imager's FOV rather locally generated and evolved within the FOV. The generation of the EMSTIDs is well understood through the Perkins instability mechanism in combination with some seeding mechanisms such as E-F coupling, interhemispheric coupling, and gravity wave (Cosgrove and Tsunoda, 2004;

Huang et al., 1994; Kelley & Makela, 2001; Perkins, 1973; Otsuka et al., 2004; Shiokawa et al., 2005). Therefore, the exact seeding source for the observed EMSTID's generation is not the focus of the present communication. It rather focuses on the role of prompt electric fields on the evolution and decay of the EMSTID during two consecutive substorms.

As eastward and westward electric fields were imposed over the Hanle region during the two substorms, therefore, here we have explored the role of these electric fields on the evolution and decay of the observed EMSTID and the probable physical processes provided in the schematic diagram shown in Figure 5. Figure 5a on the left panel shows the quiet time background conditions in the nighttime mid-latitude F-region. Here, the direction of neutral wind ($U$) is in southeast (SE) direction (Hamza et al., 1999; Shiokawa et al., 2003b; Sun et al., 2015) represented by blue dashed arrow and the background electric field ($E$) is in southwest (SW) direction (Otsuka et al., 2004; Sun et al., 2015; and references therein) represented by orange arrow. These two parameters ($U$ and $E$) are responsible for generating a current $J = \Sigma(E + U \times B)$, where $\Sigma$ is the field line integrated Pedersen conductivity. It is observed that the magnitude of the nighttime background electric field ($E$) is comparatively less, therefore $U \times B$ contributes to the $J$, which will be in the northeast (NE) direction. This north-eastward current generates polarization electric fields ($\pm\delta E$), yellow arrows represent the directions of these electric fields shown in the left panel of Figure 5b. These oppositely directed (NE and SW) polarization electric fields cause undulations in the plasma density and hence generates an initial perturbation. The NE polarization electric field generates plasma depleted regions (blue fronts), while SW polarization electric field generates plasma enhanced regions (red front) through vertical $\delta E \times B$ drift (shown in Figure 5b). The density fluctuations corresponding to the plasma depleted and rich regions are represented by the blue and red curves, respectively. The directions of the vertical drifts are shown by red arrows, while the zonal components of the polarization electric fields are shown by black arrows. If the initial perturbation has sufficient growth rate, it can grow and generate MSTIDs. The subfigures on the right panel represent the plasma density variations during initial perturbation (Figure 5c), after the external eastward (Figure 5d), and westward (Figure 5e) electric field, respectively. Considering $+\delta E$ as the eastward component of the polarization electric field within the plasma depleted region and $-\delta E$ as the westward component of the polarization electric field within the plasma rich region of the initial density perturbation. When the external eastward electric field imposed over the Hanle region during the first substorm, it enhanced growth rate of the Perkins instability via increasing the polarization electric field (due to the enhancement of background current ($J$)) and reducing the ion-neutral collision frequency (due to ascent of the F-layer shown in Figures 2 & 3b). Hence, during the first substorm, eastward imposed electric field increased the growth rate of the Perkins instability and assisted the evolution of the EMSTID (evident from the increase in the EMSTID's amplitude as shown in Figures 3c & 5d). Whereas during the second substorm, the westward electric field decreased the growth rate of the Perkins instability via decrease in the background current ($J$) and increase in the ion-neutral collision frequency and consequently led to the decay of the EMSTID (amplitude decrease) as shown in Figures 3c & 5e. However, we have observed that the EMSTID amplitude does not show instantaneous effect to the second substorm. The response of the EMSTID amplitude was delayed by ~19 minutes. The reason is due to the interplay between the polarization electric field and the externally induced westward electric field. The reduction in the amplitude increment rate after the onset of the second substorm up to 15.2 UT and the decrease in the amplitude afterwards (shown by blue curve in Figure 3c) shows clear evidence of weakening of the polarization electric field associated with EMSTID front. The polarization electric field ($\delta E$) associated with the MSTID bands is related to the F-region height integrated Pedersen conductivity ($\sum_P^F$) and effective electric field as (Otsuka et al., 2007)

$$\boldsymbol{\delta E} = \frac{\delta \sum_P^F}{\sum_P^F}(\boldsymbol{E} + \boldsymbol{U} \times \boldsymbol{B}) \cdot \frac{\boldsymbol{k}}{|\boldsymbol{k}|} \qquad (4)$$

where, $\boldsymbol{\delta \sum_P^F}$ is the perturbation of $\sum_P^F$ from the background, $\boldsymbol{E}$ is the background electric field, $\boldsymbol{U}$ is the neutral wind, and $\boldsymbol{B}$ is the geomagnetic field. It is evident from Equation (4) that the polarization electric field decreases (increases), if the background Pedersen conductivity $\sum_P^F$ increases (decreases) and/or effective electric field ($\boldsymbol{E} + \boldsymbol{U} \times \boldsymbol{B}$) decreases (increases). It is evident from Figure S2 (shown in supplementary document) that $\sum_P^F$ increased (also evident form the enhancement of the foF2 between 15 and 15.5 UT; shown in Figure 3b) between 14.75 and 15.5 UT due to the passage of enhanced plasma region over the observation region. Most importantly, after the second substorm onset, externally imposed westward electric field reduced the effective electric field. Therefore, during the second substorm, these two factors weakened the polarization electric field (shown in Equation 4), which eventually reduced the EMSTID amplitude and led to the decay of the EMSTID.

Along with the evolution and decay of the EMSTID, we also observed a new NW-SE oriented plasma depleted front at ~14.87 UT (20.37 IST), which later split into three thin fronts and eventually disappeared. This new front might have been formed as part of the existing EMSTID or due to another structure that existed at higher altitude region. During the second substorm, due to the downward $\boldsymbol{E} \times \boldsymbol{B}$ drift the structure descended to the 630.0 nm airglow altitude range and was captured in the airglow images. The existence and splitting of this new front (marked by cyan colored lines in Figure 1h-l) are beyond the scope of this paper rather it mainly focusses on the role of the externally imposed electric fields during the two consecutive substorms on the evolution and decay of the EMSTID over Hanle region. During the first substorm, the eastward electric field perturbations might have assisted the evolution of the EMSTID whereas, the westward prompt electric field perturbations during the second substorm led to the decay of the EMSTID. This study provides evidence for the evolution and decay of EMSTIDs during substorm conditions for the first time. Thus, it also gives a hint that for the better prediction of the ionospheric variabilities associated with EMSTIDs over the mid-latitudes one should consider the storm and substorm effects as well.

## 5. Summary and Conclusion

This study elicits the effects of interplanetary electric field ($IEF_y$) and substorm induced prompt electric fields on the mid-latitude ionospheric plasma structures over Hanle region in India. On a moderately geomagnetic active night of 26 October 2019, we observed the evolution and decay of an EMSTID during two consecutive substorms. The main conclusions of this study are as follows:

1. The h'F enhancement and the reduction in the intensity of the $O(^1D)$ 630.0 nm airglow images with evolution of the EMSTID between 13.52-14.56 UT (19.02-20.06 IST) confirmed the existence of an eastward electric field over Hanle region. There exist two possibilities for eastward electric field perturbation over Hanle during the first substorm. One, both prompt penetration of $IEF_y$ and the substorm induced prompt electric field were eastward and contributed to the net eastward electric field perturbation. Two, the substorm induced prompt electric field was westward and the eastward prompt penetration of $IEF_y$ overpowered the substorm induced prompt electric field generating a net eastward electric field perturbation. This eastward electric field perturbation assisted the evolution of the EMSTID.

2. The reduction in h'F and the enhancement in the intensity of the $O(^1D)$ 630.0 nm airglow images with decay of the EMSTID between 14.87-15.8 UT (20.37-21.3 IST) confirmed the existence of a westward electric field. It appears that during the second substorm, westward prompt electric field perturbations caused by the substorm contributed to the net westward electric field perturbations. This westward electric field led to the decay of the EMSTID in the later stage.

**Data Availability Statement**

The all-sky airglow unwarped images and the entire ionograms on the night of 26 October 2019 are available at Rathi et al. (2024b). Geomagnetic activity indices (Ap and AL) were obtained from the World Data Center (WDC), Kyoto (https://wdc.kugi.kyoto-u.ac.jp/wdc/Sec3.html). The neutral wind data was obtained from Horizontal Wind Model (HWM14). NRL-MSISE00 model data is available at https://ccmc.gsfc.nasa.gov/modelweb/models/msis_vitmo.php. The interplanetary electric field data is obtained from NASA GSFC CDAWeb (https://cdaweb.gsfc.nasa.gov/index.html). ASY-H (asymmetric component of ring current) index, SYM-H (symmetric component of ring current) index, PC (polar cap) index, and GOES electron fluxes obtained from NASA GSFC CDAWeb are also used in this study. The wave and planetary (Wp) index data is obtained from https://www.isee.nagoya-u.ac.jp/~nose.masahito/s-cubed/. SMU and SML indices are obtained from SuperMAG (https://supermag.jhuapl.edu/), which are derived using more than 100 ground magnetometers data (Gjerloev, 2012; Newell & Gjerloev, 2011).

**Acknowledgements**

S. Sarkhel acknowledges the financial support from the Science and Engineering Research Board, Department of Science and Technology, Government of India (CRG/2021/002052) to maintain the multi-wavelength airglow imager. The support from Indian Astronomical Observatory (operated by Indian Institute of Astrophysics, Bengaluru, India), Hanle, Ladakh, India for the day-to-day operation of the imager is duly acknowledged. We gratefully acknowledge the SuperMAG collaborators (https://supermag.jhuapl.edu/info/?page=acknowledgement). We are also thankful to the GOES team for making their data available to the public. R. Rathi acknowledges the fellowship from the Innovation in Science Pursuit for Inspired Research (INSPIRE) programme, Department of Science and Technology, Government of India. M. Sivakandan acknowledges the Alexander Von Humboldt Foundation and Leibniz Institute of Atmospheric Physics (IAP) for providing the postdoctoral fellowship. The work of D. Chakrabarty is supported by the Department of Space, Government of India. This work is also supported by the Ministry of Education, Government of India.

**Figures**

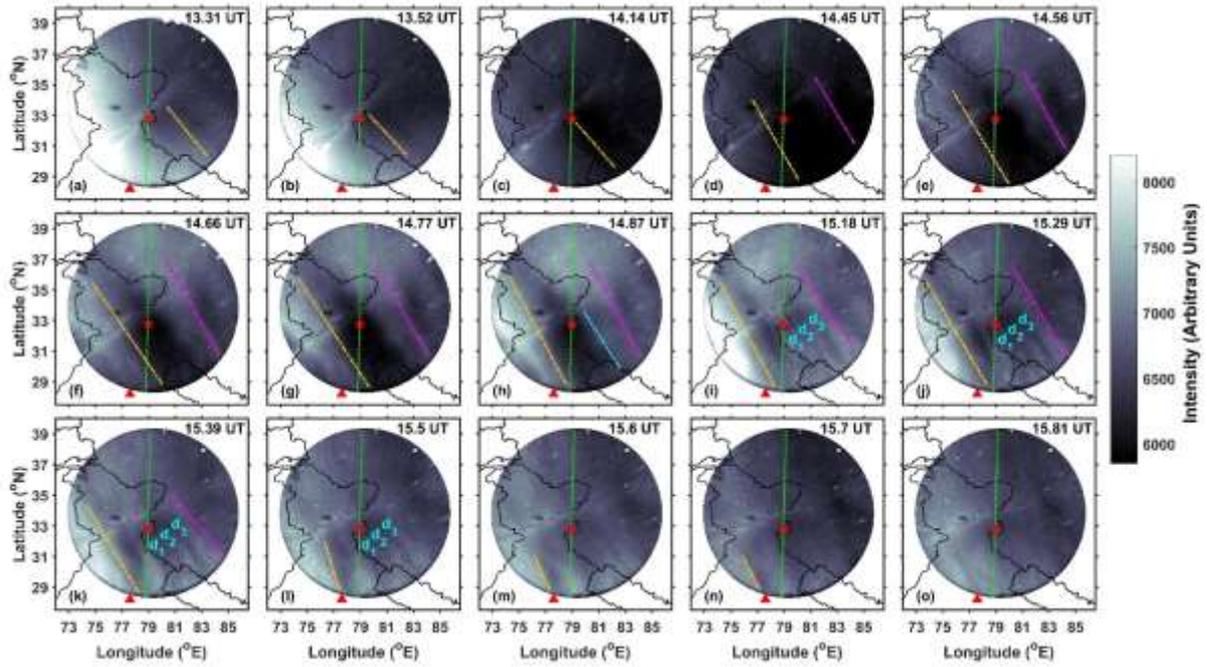

**Figure 1:** A sequence of the images at 630.0 nm, showing the evolution, dissipation of EMSTID and splitting of a plasma depleted front on the night of 26 October 2019 using an all-sky airglow imager located at Hanle, Ladakh, India. The red star and triangle represent the locations of the imager at Hanle and the digisonde at New Delhi, India, respectively. The nearly north-south aligned green dashed-dotted line represents the orientation of geomagnetic field lines in the ionosphere over Hanle. All the plasma depleted fronts are marked with different colored dashed-dotted lines and the thin fronts after the splitting are marked as '$d_1$', '$d_2$', and '$d_3$'.

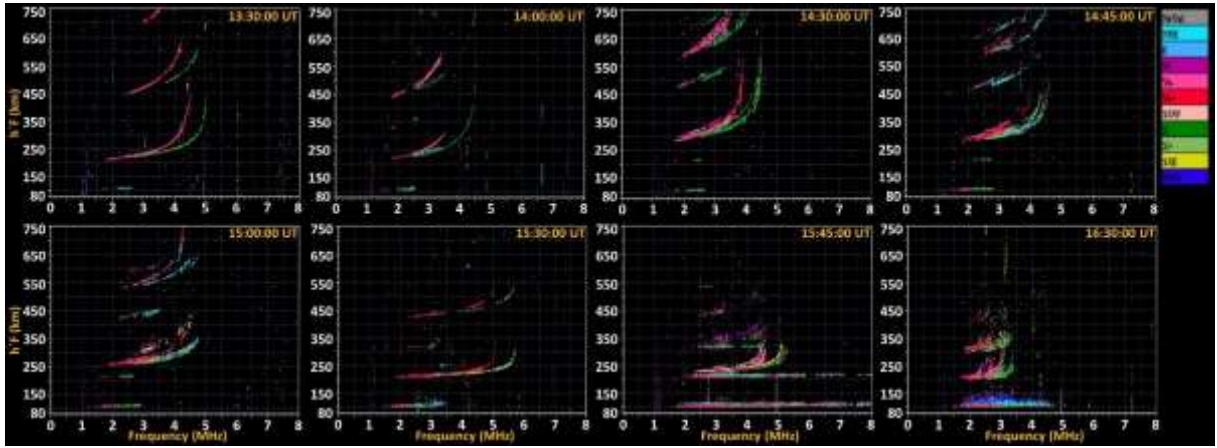

**Figure 2:** A sequence of ionograms on the night of 26 October 2019 from New Delhi digisonde showing rising/falling of F-layer trace and presence of spread F.

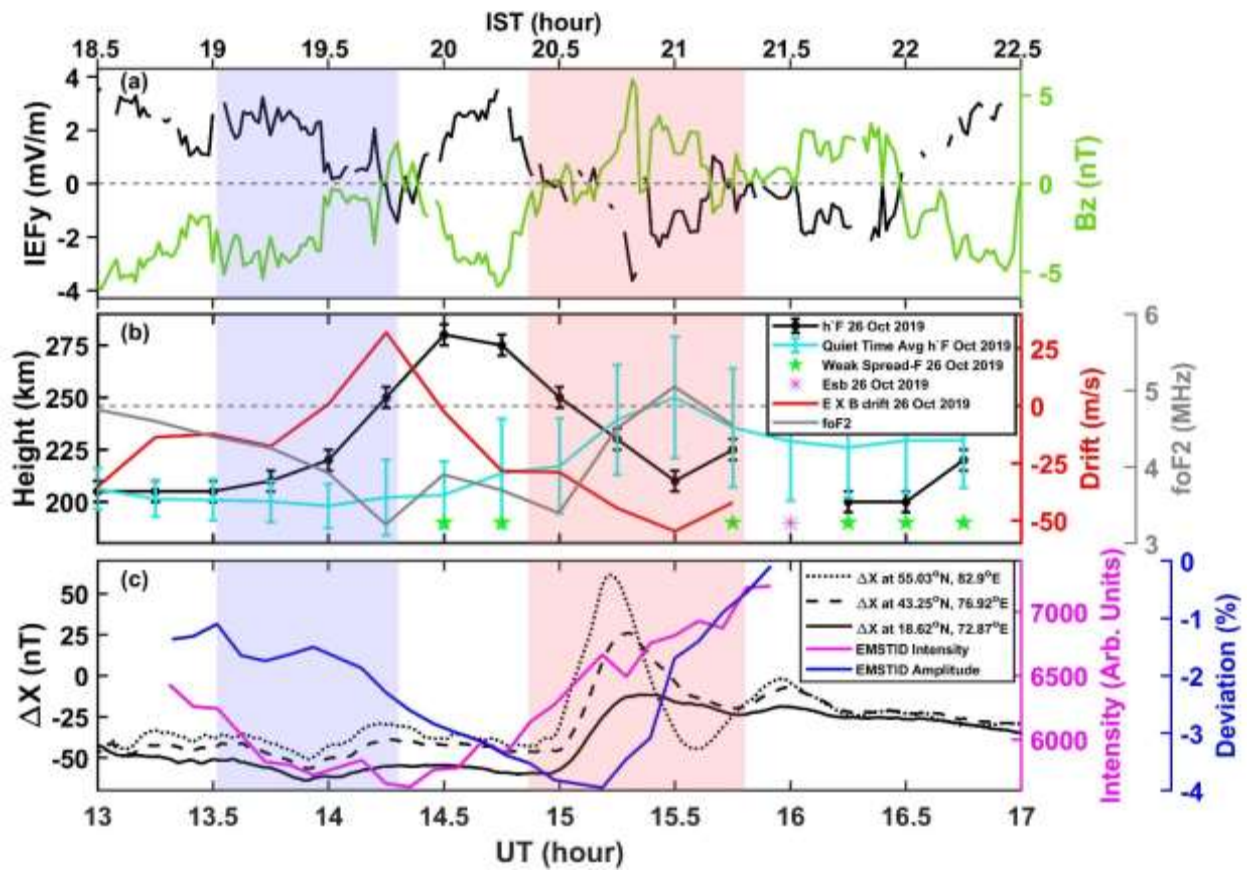

**Figure 3:** (a) shows variation of *Y*-component of interplanetary electric field (IEF$_y$) and interplanetary magnetic field (IMF $B_z$). (b) shows h'F variation over New Delhi on 26 October 2019 (event day) and quiet time average h'F of October month. The red and gray curve show variation of **E** × **B** drift and foF2 on 26 October 2019, respectively. Green stars and magenta asterisks are showing the presence of weak spread-F and blanketing sporadic E (Esb) layer on 26 October 2019. (c) shows the variation of the change in the north-south component (ΔX) of the geomagnetic field from three ground magnetometers near to the Hanle longitude (78.9°E). The magenta and blue curve show EMSTID's intensity with background contamination and EMSTID amplitude excluding the background on 26 October 2019. The blue and red shaded regions show the period of the two substorms.

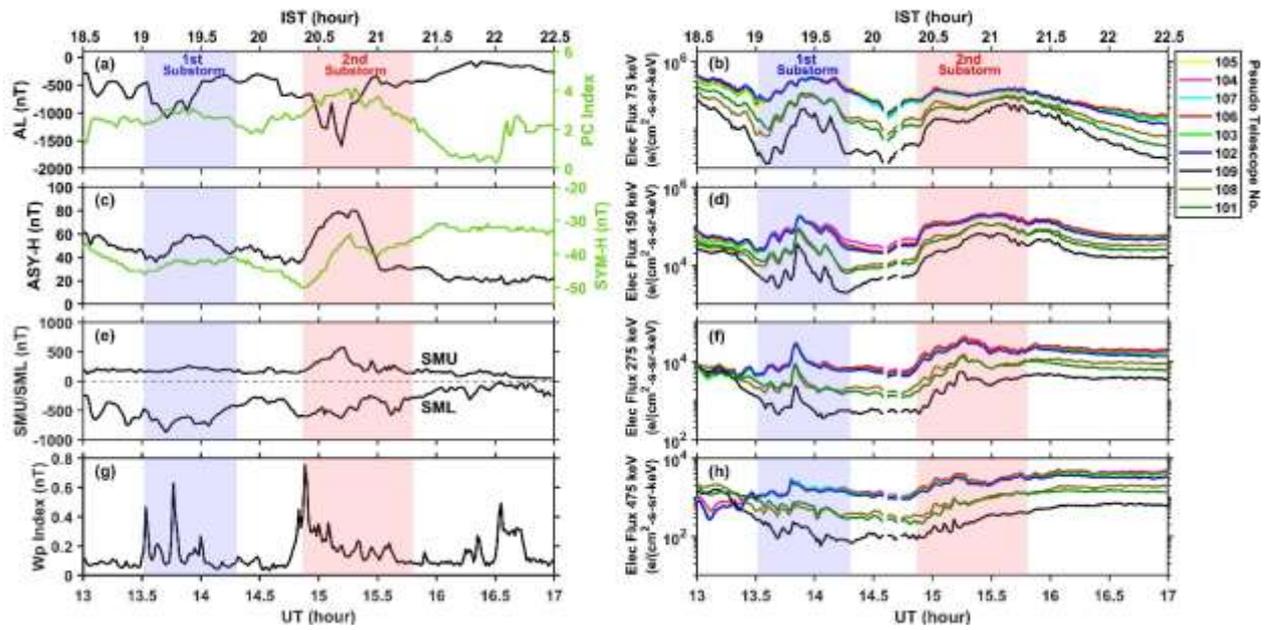

**Figure 4:** (a, c, e, & g) show different geomagnetic indices viz. AL, PC, ASY/H, SYM/H, SMU, SML, & Wp. (b, d, f, & h) show the GOES electron fluxes at 75, 150, 275, & 475 keV. In all the subfigures the blue and red shaded regions represent the duration of the first and second substorm, respectively.

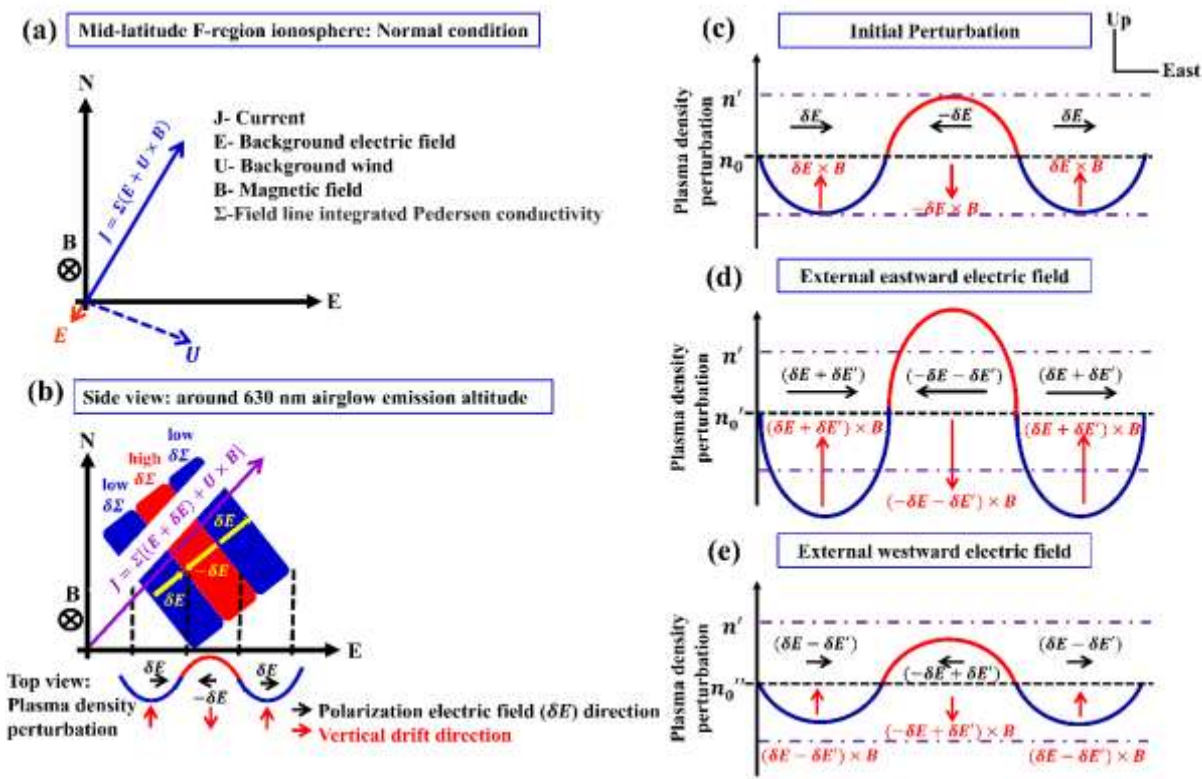

**Figure 5:** Schematic diagram shows the different conditions of mid-latitude ionospheric F-region. (a & b) on the left panel show the normal background conditions and initial perturbation for Perkins instability. (c, d, & e) on the right panel show the initial plasma density perturbation, plasma density perturbation after the substorm induced eastward electric field, and plasma density perturbation after the substorm induced westward electric field. The blue horizontal dash-dotted lines represent the plasma density variation during the initial perturbation. Black horizontal arrows represent the directions of the electric fields, while vertical red arrows represent the direction as well as magnitude of the vertical plasma drifts.